\documentclass[prl,twocolumn,showpacs,floatfix]{revtex4}
\usepackage{amsmath} %some extra symbols
\usepackage{amsfonts} %some extra symbols
\usepackage{amssymb} %some extra symbols
\usepackage{bm} %allows bold italic
\usepackage{graphicx}% needed for figures
\usepackage[applemac]{inputenc}
\usepackage{hyperref}
\usepackage{epstopdf} %muuttaa automaattisesti kuvat pdf-tiedostoiksi
\begin{document}
\title{Pendulum in Fermi liquid}
\author{Timo H. Virtanen}
%\email[]{}
\affiliation{Department of Physics, University of Oulu, Finland}
\author{Erkki Thuneberg}
%\homepage[]{Your web page}
%\thanks{}
\affiliation{Department of Physics, University of Oulu, Finland}
\date{\today}
\begin{abstract}
The Fermi liquid theory formulated by Landau is a basic paradigm of the behavior of an interacting many-body system. We present a new application of this theory to calculate "Landau force" on a macroscopic object.
We show that immersing a pendulum in Fermi liquid can increase its oscillation frequency, and evidence of this has been observed in mixtures of $^3$He and $^4$He. 
\end{abstract}
\pacs{67.10.Db, 62.10.+s, 67.60.G-}
%67.60.G- Solutions of 3He in liquid 4He
%62.10.+s	Mechanical properties of liquids
%67.10.Db	Fermion degeneracy
 
\maketitle
%\tableofcontents

Pendulum is one of the simplest but also most precise physical instruments. For accurate measurements, one needs to understand the effect of the medium, often air, on the oscillation. The first effect perceived was to correct the gravitational restoring force by the buoyancy of the oscillating body. The second effect, calculated by Poisson in 1831, was that the flow of the medium around the body increases the inertia of the pendulum. The third correction, calculated by Stokes in 1850, was the effect of the viscosity of the medium \cite{Stokes}. With these corrections, the pendulum in fact turns out to be a device, a viscometer, to measure the properties of the medium. Further corrections are needed in a rarified gas, where the mean free path of the gas particles becomes comparable or exceeds the size of the pendulum \cite{Knudsen33,Cercignani68}.
In this article we continue this line of corrections by considering a pendulum in Fermi liquid. We show that the Fermi liquid acts like an elastic medium, which can increase  the restoring force and thus lead to increased frequency of oscillation. We show that this effect is important in explaining oscillator experiments made in liquid mixtures of $^3$He and $^4$He [\onlinecite{Martikainen02,Pentti09}].  Moreover, this fermion-boson mixture allows a deep insight into the  Fermi-liquid theory.

The Fermi-liquid theory is a paradigm of what can happen in a strongly interacting many-particle system
 \cite{Landau57,LLs2,BP,Shankar94}. Instead of strongly interacting particles, the energy spectrum has low-energy excitations called quasiparticles, which interact only weakly. Originally Landau formulated the Fermi-liquid theory for liquid $^3$He [\onlinecite{Landau57}], but the principle has much wider application,  an important example being the conduction electrons in metals. The generalization of Fermi-liquid theory forms the basis to understand the superconductivity of many materials and the superfluidity in fermion systems. Even when it is not valid, it forms the standard against which to compare more sophisticated theories \cite{Shankar94,Jain09}.

In the original realm of $^3$He, the Fermi-liquid theory was applied to explain many properties of the bulk, in particular thermodynamics, collective modes and transport properties \cite{LLs2,BP}. Soon the theory was applied to a liquid limited by an oscillating planar wall \cite{Bekarevich61,Flowers78}. This problem still had an analytic solution although it is very complicated, and therefore also approximate methods are useful \cite{Richardson78}. The approaches to non-planar geometries have been perturbative. As the Fermi-liquid theory reduces to the hydrodynamic theory in the limit of small mean free path, the leading corrections appear in the boundary conditions. This leads to a modified boundary condition where the fluid velocity is assumed to extrapolate to the wall velocity at a ``slip length'' behind the wall \cite{HojgaardJenssen80,Carless83,EinzelParpia97,Bowley04,Perisanu06}. In this letter we calculate the response of the Fermi liquid to a vibrating body in the full range of mean free paths from the hydrodynamic to the ballistic limit. The calculation necessarily is numeric. With a view to apply the calculation to measurements using vibrating wires, we specifically consider a circular cylinder oscillating transverse to its axis.
We also use the Fermi-liquid theory adapted to the simultaneous presence of bosonic superfluid, as formulated by Khalatnikov \cite{Khalatnikov69}. Besides being applicable to mixtures of $^3$He and $^4$He, this has the advantage that some basic properties of the Fermi-liquid theory are more easily visible \cite{perusyht}.

A noninteracting Fermi system has plane-wave states with momenta $\bm p$ and energy $\epsilon_{\bm p}=p^2/2m$ where $m$ 
is the mass of a fermion. The ground state of the system consists of a Fermi sphere with all states $p<p_{\rm F}$ filled and others empty.
The basic assumption of Landau was that the energy spectrum of the interacting Fermi system has one-to-one correspondence with the non-interacting system. In particular, the momenta $\bm p$ of the quasiparticles are the same as for noninteracting particles, but the energies are shifted. Near the Fermi surface one writes the energy as linear in momentum, $\epsilon_{\bm p}=v_{\rm F}(p-p_{\rm F})$ where the parameter $v_{\rm F}=p_{\rm F}/m^*$ defines the effective mass $m^*$.
Using the group-velocity argument to the dispersion relation $\epsilon_{\bm p}$ gives that the excitation propagates with velocity $\bm v$ whose magnitude equals the Fermi velocity $v_{\rm F}$. Thus the momentum of the principal fermion $m\bm v$ differs from the momentum $\bm p$ of the excitation. In Fermi-liquid theory this missing momentum is parameterized so that fraction $D=1-(1+\frac13F_1)m/m^*$ of the total momentum $\bm p$ is carried by the bosons and fraction $mF_1/3m^*$ by other fermions. This can be interpreted that the principal fermion pushes with it a cloud of both bosons and other fermions.
Here the bosons are assumed to be fully condensed and are described by density $\rho_{\rm B}$, chemical potential $\mu_{\rm B}$ and superfluid velocity $\bm v_s$. The special case without bosons is obtained by setting $\rho_{\rm B}=D=0$. 

A crucial observation of Landau was that the dispersion relation assumed above leads to a consistent theory  only if one allows an interaction between the quasiparticles. This modifies the quasiparticle energy to $\epsilon_{\bm p}=v_{\rm F}(p-p_{\rm F})+\delta\epsilon_{\hat{\bm p}}$ with the correction \cite{Khalatnikov69}
\begin{eqnarray}
\delta\epsilon_{\hat{\bm p}}({\bm r},t)=(1+\alpha)\delta\mu_{\rm B}({\bm r},t)+
Dp_{\rm F}\hat{\bm p}\cdot{\bm v}_s({\bm r},t)\nonumber\\+
 \sum_{l=0}^\infty
F_l\langle P_l(\hat{\bm p}\cdot\hat{\bm p}')\phi_{\hat{\bm p}'}({\bm
r},t)\rangle_{\hat{\bm p}'}.
\label{e.deltae10}\end{eqnarray}
Close to the Fermi surface, $\delta\epsilon_{\hat{\bm p}}$ depends on momentum only through its direction $\hat{\bm p}=\bm p/p$. The first term in Eq.\ (\ref{e.deltae10}) is the energy shift  due to a non-equilibrium boson chemical potential $\delta\mu_{\rm B}=\mu_{\rm B}-\mu_{\rm B}^{(0)}$, parameterized by $\alpha$. The second term is the energy shift   due to motion of the boson part with velocity $\bm v_s$.  In the last term  $\phi_{\hat{\bm p}}$ is the energy shift of the Fermi surface, $P_0(x)=1$, $P_1(x)=x$,\ldots  are the Legendre polynomials and $\langle \ldots\rangle_{\hat{\bm p}}$ is the average over the Fermi surface. The terms with coefficients $F_0$ and $F_1$ are analogous to the $1+\alpha$ and $D$ terms but correspond to the fermion background. The higher order terms with $l>1$ extend the leading two terms to general deformations of the Fermi surface.

In order to determine the dynamics, one needs a kinetic equation. For small deviations from equilibrium it takes the form
\begin{eqnarray}
\frac{\partial\phi_{\hat{\bm p}}}{\partial t}
+v_{\rm F}\hat{\bm p}\cdot\bm\nabla(\phi_{\hat{\bm p}}+\delta\epsilon_{\hat{\bm p}})=I,
\label{e.lb}\end{eqnarray}
where $I$ is the collision term. We see that this equation is a first order differential equation along classical particle trajectories, which are straight lines in the momentum direction $\hat{\bm p}$. The momentum conservation allows to determine the stress tensor and thus the force on macroscopic objects.

\begin{figure}[tb] %  figure placement: top, bottom, (not here or page)
   \centering
   \includegraphics[width=0.60\linewidth]{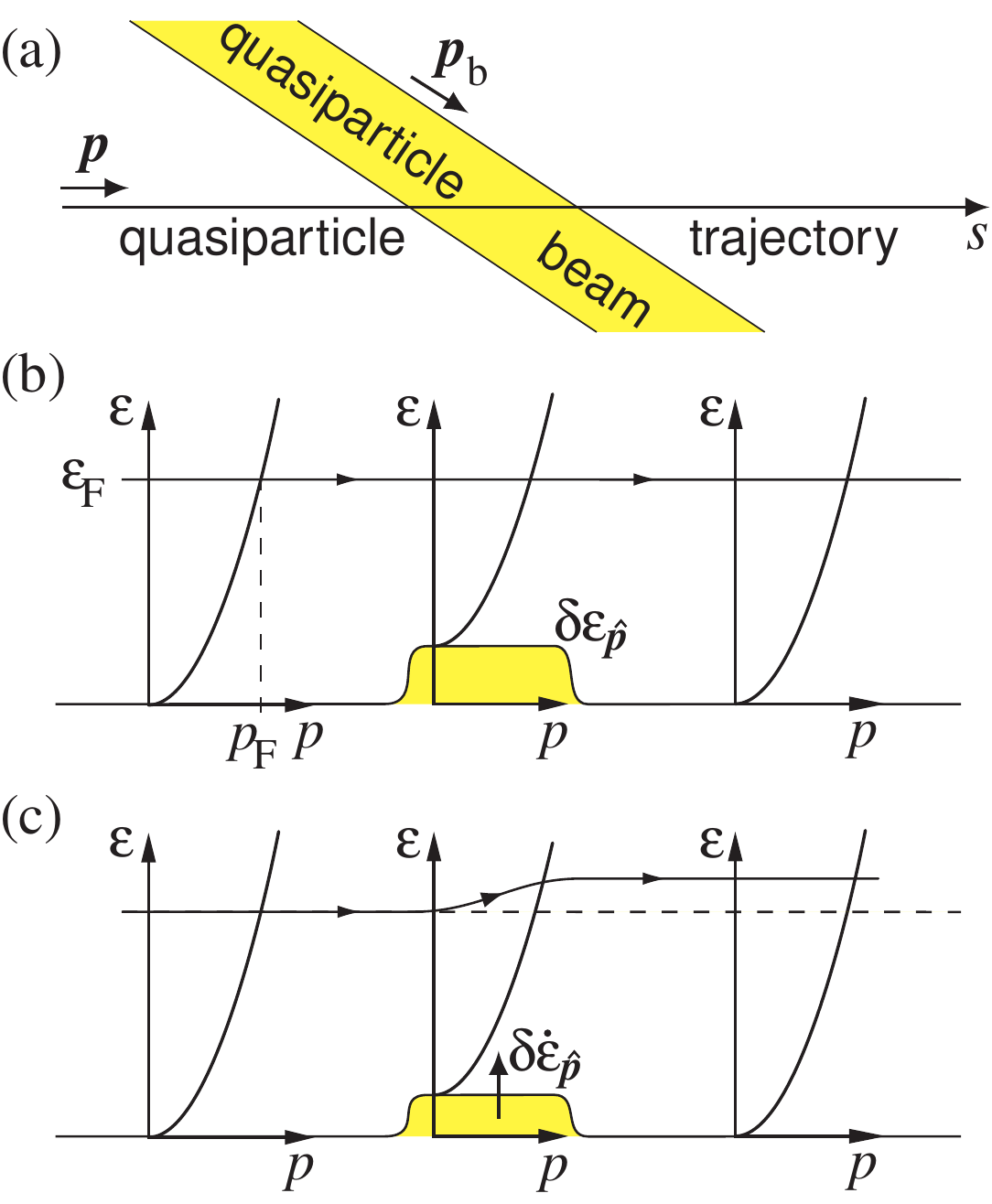}
 \caption{Illustration of quasiparticle dynamics on a trajectory crossing a beam of quasiparticles. (a) The quasiparticle trajectory  (with parameter $s$) and  the beam depicted in $\bm r$ space. (b)  The quasiparticle dispersion relations $\epsilon_{\bm p}$ at three locations on the quasiparticle trajectory crossing the beam. For a stationary $\delta\epsilon_{\hat{\bm p}}$ the quasiparticles travel at constant energy and are not essentially affected by the beam. (c) A temporal variation of the  potential ($\delta\dot\epsilon_{\hat{\bm p}}\not=0$) leads to energy shift of quasiparticles. }
   \label{f.qpbeam}
\end{figure}

In order to see how the elasticity of the liquid arises, let us consider a beam of quasiparticles, as depicted in Fig.\  \ref{f.qpbeam}. On a trajectory crossing the beam, the beam causes a potential $\delta\epsilon_{\hat{\bm p}}$, which shifts all quasiparticle energies (Fig.\  \ref{f.qpbeam}b).  As in equilibrium the states are filled up to the Fermi level, the potential is compensated by a change of  the particle density on the crossing trajectory.  Therefore, the dynamics on the crossing trajectory, which is determined by quasiparticles at energies close to the Fermi level, is not essentially affected by a stationary beam.
%%%%%%%%%%%%%%%%%%%%%%%%%%%%%%%%%%%%%%%%%%%
%%%%%%%%%%%%%%%%%%%%%%%%%%%%%%%%%%%%%%%%%%%

Let us now consider that the intensity of the beam is changing. The changing of the potential stores or releases quasiparticles on the crossing trajectory. Thus a varying quasiparticle beam radiates quasiparticles in all directions even in the absence of any collisions ($I=0$). In the case of a pendulum,  the  body moving with velocity $\bm u$  generates in the radial direction $\hat{\bm r}$ a beam with amplitude proportional to  $\bm u\cdot\hat{\bm r}$.  Typically  the $F_0$ term is the dominant interaction term in (\ref{e.deltae10}) and thus  the potential $\delta\epsilon_{\hat{\bm p}}\propto F_0\bm u\cdot\hat{\bm r}$. The quasiparticles radiated back on the body are proportional to $\delta\dot\epsilon_{\hat{\bm p}}\propto F_0\dot{\bm u}\cdot\hat{\bm r}$, where the dot denotes time derivative. This results in an extra ``Landau force'' $\bm F\propto  F_0\dot{\bm u}$ the body has to exert on the liquid.
A force of the form $\bm F=M\dot{\bm u}$ is well known in hydrodynamic flow, and $M$ can be interpreted as the mass of the fluid that is dragged with the body. The difference in the present case is that $F_0$ can be negative. Rather than thinking of a negative mass, a simpler interpretation is that the fluid has elasticity leading to an increased restoring force and increased oscillation frequency of the pendulum.

In pure $^3$He  the parameter $F_0$ is positive. The Landau force leads to increased effective mass of any objects, including ions \cite{Borghesani}, aerogel \cite{aerogel} and deformations of the surface of the liquid \cite{Kono08}, which effect has not been explored yet.
The case of negative $F_0$ is realized in mixtures of $^3$He and $^4$He. Below we calculate in detail the case of a circular cylinder oscillating transverse to its axis.

\begin{figure}[t!] %  figure placement: top, bottom, (not here or page)
   \centering
   \includegraphics[width=0.70\linewidth]{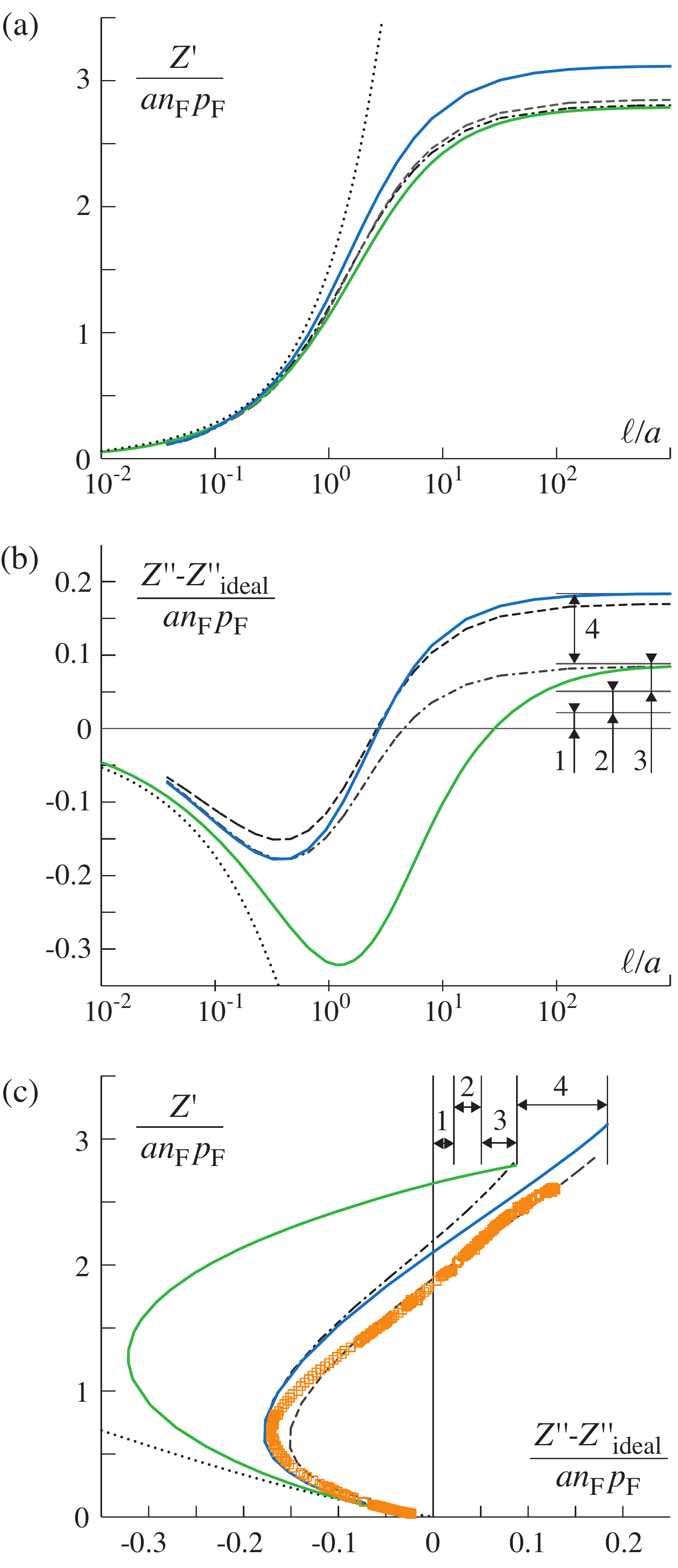}
  \caption{The force $\bm F=Z\bm u$ exerted by an oscillating cylinder  (radius $a$, velocity $\bm u$) on the surrounding Fermi-Bose liquid as a function of the mean free path $\ell$. (a) The real part $Z'$ and (b) the imaginary part $Z''$ of the mechanical impedance $Z$. The main results are shown by blue solid lines. They correspond to diffusive boundary conditions in a slab chamber, which has two plane walls at $x=\pm 8 a$ and the cylinder in the middle oscillating in the $x$  direction. 
Other lines differ from the blue ones in the following respects: (green solid lines) a large chamber, (dotted lines) hydrodynamic approximation in a large chamber  \cite{Stokes,Carless83}, (dashed lines) 50 \% specular scattering at the oscillating cylinder, and (dash-dotted lines) quasiparticle-absorbing chamber walls.  Four different contributions to the ballistic limit of the blue curve are indicated (see text). (c) Plot of $Z'$ vs. $Z''$ showing the same results as above and  experimental data (orange)
from Ref.\ \onlinecite{Martikainen02} at 5.6\% $^3$He concentration. The parameters of the calculation  $\omega a/v_{\rm F}=0.017$,  $F_0=-0.28$ and $F_1=0.155$ were fixed by independent measurements. The large chamber is a coaxial cylinder of radius $b=80a$ with absorbing walls.  
 }
   \label{f.zmfp}
\end{figure}

The system of equations  (\ref{e.deltae10})-(\ref{e.lb}) was solved numerically for $\phi_{\hat{\bm p}}(\bm r)$  on a grid surrounding an oscillating cylinder using relaxation-time approximation for the collision term.
The force $\bm F$ by which the cylinder drives the liquid is conveniently expressed by mechanical impedance $Z=Z'+iZ''$ defined by $\bm F=Z\bm u$, where  we assume time dependence $\exp(-i\omega t)$. The dissipative $Z'$ and the reactive $Z''$ are plotted in
Fig.\ \ref{f.zmfp} as a function of the mean free path $\ell$. Two cases are shown by solid lines: a cylinder of radius $a$ in a large chamber and in a slab of thickness $16a$. We see that the confinement has strong effect especially on $Z''$.
%%%%%%%%%%%%%%%%%%%%%%%%%%%%%%%%%%%%%%%%%%%
%%%%%%%%%%%%%%%%%%%%%%%%%%%%%%%%%%%%%%%%%%%

In the limit $\ell\rightarrow 0$, the viscosity can be neglected and the liquid behaves like an ideal fluid. This corresponds to pure reactance, $Z_{\rm ideal}=-i\pi a^2\rho \omega G$. Here $\rho$ is the liquid density and $G-1=O(a/b)^2$  is a small correction caused by a finite chamber dimension $b$. In the experimental case \cite{Martikainen02} $Z''_{\rm ideal}/an_{\rm F}p_{\rm F}=-0.50$. With increasing $\ell$ the viscosity of the liquid becomes important. This leads to increasing dissipation and the effective mass $M=-Z''/\omega$ grows because the increasing viscosity causes  more liquid to be dragged by the oscillating cylinder.
The hydrodynamic description becomes insufficient when $\ell$ approaches $a$. Compared to the hydrodynamic theory (dotted lines), the dissipation grows less rapidly and the fluid mass coupled to the oscillator starts to decrease. In the ballistic regime $\ell\gg a$ the impedance saturates to a value independent of $\ell$.

The fact that $Z''$ in the ballistic limit  exceeds its ideal-fluid value in the limit $\ell\rightarrow0$ can be analyzed as follows.
1) The fermion component of the fluid decouples from the ideal fluid flow corresponding to contribution $mn_{\rm F}$ to be subtracted from $\rho$ in $Z_{\rm ideal}$. (Here $n_{\rm F}$ the fermion number density and $m$ the bare fermion mass.)
2)  As discussed above, part of the bosons are bound to fermion quasiparticles, and therefore the density  $Dm^*n_{\rm F}/(1+\frac13F_1)$ has to be subtracted from $\rho$. These two contributions together constitute what is known as the ``normal fluid density''  $\rho_n=m^*n_{\rm F}/(1+\frac13F_1)$. 3) The Landau force, the main contribution coming from the negative $F_0$.
4) In a finite geometry, there is an effect caused by quasiparticles reflected from the chamber wall back to the oscillating body.
All these effects are shown separately in Fig.\ \ref{f.zmfp}.
We see that the four effects are all in the same direction, and have similar orders of magnitude.

Fig.\ \ref{f.zmfp}c gives a comparison of the theory to experiments \cite{Martikainen02,Pentti09}. Experimentally the mean free path was controlled by temperature,  the lowest temperatures approaching the ballistic limit.
We see that the confinement is crucial in order to achieve agreement. The essential point in the data is that the frequency shift of the oscillator in the low temperature limit is positive and clearly larger than the normal fluid contribution (contributions 1 and 2 together). This indicates the importance of the contributions of Fermi-liquid interactions and confinement, although their individual contributions cannot be separated.

In the calculation,  the parameters of the slab chamber and the oscillating cylinder are fixed by the experiment. The Fermi-liquid parameters $m^*=2.46$,  $F_0=-0.28$ and $F_1=0.155$ are based on measurements of specific heat and second sound velocity  as analyzed by Corruccini  \cite{Corruccini84}. The higher Landau parameters $F_l$ with  $l>1$ were assumed to vanish. The boundary condition corresponding  to diffuse scattering of quasiparticles \cite{Bekarevich61} was used on all surfaces for the blue solid lines in Fig.\ \ref{f.zmfp}. Thus no fitting parameters was used. 

In more detailed comparison, one can allow for partial specularity of the quasiparticle scattering from the  surface of the oscillating cylinder \cite{Perisanu06}. Also, the walls of the experimental chamber are rough sinter, where part of the quasiparticles likely is absorbed rather than scattered.  The effect of these modifications are indicated by dashed and dash-dotted lines in Fig.\ \ref{f.zmfp}. The effect of changing the other parameters and full comparison with the experiments \cite{Martikainen02,Pentti09}, which were made using two different wires and at different concentrations and pressures, will be presented elsewhere \cite{long}.

A Fermi liquid can support a collisionless propagating mode, known as zero sound. The zero sound arises from the same physics as the Landau force discussed here [Eqs.\ (\ref{e.deltae10})-(\ref{e.lb})]. Using terms of antenna theory, the Landau force could be interpreted as the near field effect and zero sound as the far field effect of the Fermi-liquid interactions. The zero sound occurs for repulsive interactions ($F_0>0$ in the case where $F_0$ is the dominant interaction parameter). The elasticity effect (Landau force $\bm F$  and $\dot{\bm u}$ in opposite directions)  occurs for attractive interactions 
($F_0<0$ in the case of dominant $F_0$), where zero sound is damped.

In the ballistic limit, the Landau force can be calculated by low-frequency expansion of Eqs.\ (\ref{e.deltae10})-(\ref{e.lb}). The result is
\begin{eqnarray}
Z''_{\rm Landau}= -a^2m^*n_{\rm F} \omega \sum_{l=0}^\infty\frac{C_lF_l}{1+\frac1{2l+1}F_l}.
\label{e.landauf}\end{eqnarray}
The coefficients $C_l\sim 1$ are expressed in terms of complicated integrals. The dissipative part $Z'$ in the ballistic limit has been studied in Refs.\ \onlinecite{Bowley04} and \onlinecite{VT2}.

Before numerical calculation we made a transformation that uncouples the Fermi and Bose parts in the low frequency regime \cite{perusyht}, which is a good approximation since $\omega a/v_{\rm F}\sim 0.02$. For this a new distibution function $\psi_{\hat{\bm p}}=\phi_{\hat{\bm p}}+\delta\epsilon_{\hat{\bm p}}+Dm^*\delta\mu_B/m_B(1+\frac13F_1)$ was defined. A two dimensional grid was constructed around the cylinder. On each grid point was stored the ``old''  values of the angular averages $\langle\psi_{\hat{\bm p}}\rangle_{\hat{\bm p}}$ and $\langle\hat{\bm p}\psi_{\hat{\bm p}}\rangle_{\hat{\bm p}}$, which appear in (\ref{e.deltae10}) and in $I$ (\ref{e.lb}). Then the kinetic equation was solved for $\psi_{\hat{\bm p}}(\bm r)$ by integrating along the classical trajectories, taking into account boundary conditions \cite{Bekarevich61,perusyht}. For that the values of the angular averages had to be interpolated between the grid points. By repeating this for a number of trajectories passing trough a given grid point, ``new'' values of the averages could be calculated by numerical angular integration. This was repeated for each grid point. Because of linearity of the equations, the new values of the angular averages are obtained as linear combination of the old ones. An inhomogeneous term appears in this relation because of the boundary conditions at the oscillating cylinder. A self-consistent solution was obtained by requiring the new values to be the same as the old ones. This matrix equation was solved by numerical matrix inversion.

In conclusion, we have shown that the force of a macroscopic object on a Fermi liquid has a contribution from the interactions, caused in particular  by $F_0$. With a numerical solution of the Fermi liquid equations in proper geometry we find good agreement with measurements in $^3$He-$^4$He mixtures.

We thank N. Kopnin, E. Pentti, P. Pietiläinen, M. Saarela, J. Tuoriniemi and G. Volovik for useful discussions. We thank the Academy of Finland,  the Finnish Cultural Foundation and the National Graduate School in Materials Physics for financial support.


\begin{thebibliography}{99}

\bibitem{Stokes}G. G. Stokes, {\em Mathematical and physical papers, Vol. III} (Cambridge, 1901).

\bibitem{Knudsen33}M. Knudsen, {\em The kinetic theory of gases} (Methuen, London, 1933).

\bibitem{Cercignani68} C. Cercignani, C. D. Pagani, and P. Bassanini, Phys. Fluids {\bf 11}, 1399 (1968).

\bibitem{Martikainen02} J. Martikainen, J. Tuoriniemi, T. Knuuttila, and
G. Pickett,  J. Low Temp. Phys. {\bf 126}, 139 (2002).

\bibitem{Pentti09} E. Pentti, J. Rysti, A. Salmela, A.
Sebedash, and J. Tuoriniemi, Helsinki University of Technology report TKK-KYL-020 (2009).


\bibitem{Landau57} L. D. Landau,  Sov. Phys. JETP {\bf 3}, 920 (1957).

\bibitem{LLs2}  E. M. Lifshitz and L. P.  Pitaevski\u{\i}, {\em Statistical Physics, Part 2} (Pergamon, Oxford, 1980).

\bibitem{BP} G. Baym and C. Pethick, {\em Landau Fermi-liquid theory} (Wiley, Ney York, 1991).

\bibitem{Shankar94} R. Shankar,  Rev. Mod. Phys. {\bf 66}, 129 (1994).

\bibitem{Jain09} J. K. Jain and P. W. Anderson,  Proc. Natl. Acad. Sci. {\bf 106}, 9131 (2009).

\bibitem{Bekarevich61} I. L. Bekarevich and I. M. Khalatnikov,  Sov. Phys. JETP {\bf 12}, 1187
(1961).

\bibitem{Flowers78} E. G. Flowers and R. W. Richardson,  Phys. Rev. B {\bf 17}, 1238 (1978).

\bibitem{Richardson78} R. W. Richardson,  Phys. Rev. B {\bf 18}, 6122 (1978).

\bibitem{HojgaardJenssen80} H. Højgaard Jensen, H. Smith, P. Wölfle, K. Nagai, and T. Maack Bisgaard,  J. Low Temp. Phys. {\bf 41}, 473(1980).

\bibitem{Carless83} D. C. Carless, H. E. Hall, and J. R. Hook,  J. Low
Temp. Phys. {\bf 50}, 583 (1983).

\bibitem{EinzelParpia97} D. Einzel and J. M. Parpia, J. Low Temp. Phys. {\bf 109}, 1 (1997).

\bibitem{Bowley04} R. M. Bowley and J. R. Owers-Bradley,  J. Low Temp.
Phys. {\bf 136}, 15 (2004).

\bibitem{Perisanu06} S. Perisanu and G. Vermeulen,  Phys. Rev. B {\bf  73}, 134517  (2006).

\bibitem{Khalatnikov69} I. M. Khalatnikov,   Sov. Phys. JETP {\bf 28}, 1014
(1969).

\bibitem{perusyht} E. V. Thuneberg and T. H. Virtanen, "Fermi liquid theory of Fermi-Bose mixtures", arXiv:1010.xxxx.

\bibitem{Borghesani}A. F. Borghesani, {\em Ions and electrons in liquid helium} (Oxford, 2007).

\bibitem{aerogel} W.P. Halperin and J.A. Sauls,  arXiv:cond-mat/0408593v1 (2004).

\bibitem{Kono08}  K. Kono, H. Ikegami, and Y. P. Monarkha,  J. Phys. Soc. Japan  {\bf 77}, 111004  (2008).

\bibitem{VT2} T. H. Virtanen and E. V. Thuneberg, AIP conference proceedings {\bf 850}, 113 (2006) and J. Phys. Conference Series {\bf 150}, 032115 (2009).

\bibitem{Corruccini84} L. R. Corruccini, Phys. Rev. B {\bf 30}, 3735 (1984).

\bibitem{long} T. H. Virtanen and E. V. Thuneberg, to be submitted to arXiv:cond-mat.



\end{thebibliography}
\end{document}